\documentclass[aps,twocolumn,preprintnumbers,nofootinbib,superscriptaddress]{revtex4}
\usepackage{eurosym}
\usepackage{amsmath}
\usepackage{graphicx}
\usepackage{amsfonts}
\usepackage{array}
\usepackage{amsthm}
\usepackage{bm}
\usepackage{palatino}
\usepackage{mathpazo}
\usepackage{supertabular}
\usepackage{subfig}

\usepackage[colorlinks]{hyperref}
\usepackage{subfig}
\usepackage[font=small,labelfont=bf,justification=justified,format=plain]{caption}
\usepackage[font=small,labelfont=bf,justification=justified]{caption}
\usepackage{color}
\usepackage{booktabs}
\usepackage{multirow}
\usepackage[labelfont=bf,labelsep=quad,justification=justified]{caption}
\usepackage{xcolor}
\newcommand{\be}{\begin{equation}}
\newcommand{\ee}{\end{equation}}
\newcommand{\ba}{\begin{eqnarray}}
\newcommand{\ea}{\end{eqnarray}}
\newcommand{\bal}{\begin{align}}
\newcommand{\eal}{\end{align}}
\newcommand{\lb}{\label}

\newcommand{\bw}{\begin{widetext}}
\newcommand{\ew}{\end{widetext}}

\begin{document}

\title{Connection Between the Shadow Radius and Quasinormal Modes in Rotating Spacetimes}
\author{Kimet Jusufi}
\email{kimet.jusufi@unite.edu.mk}
\affiliation{Physics Department, State University of Tetovo, Ilinden Street nn, 1200,
Tetovo, North Macedonia}
\affiliation{Institute of Physics, Faculty of Natural Sciences and Mathematics, Ss. Cyril
and Methodius University, Arhimedova 3, 1000 Skopje, North Macedonia}

\begin{abstract}
Based on the geometric-optics correspondence between the parameters of a quasinormal mode and the conserved quantities
along geodesics, we propose an equation to calculate the typical shadow radius for asymptotically flat and rotating black holes when viewed from the equatorial plane given by
\begin{equation}\notag
\bar{R}_s=\frac{\sqrt{2}}{2}\left(\sqrt{\frac{ r_0^{+}}{f'(r)|_{r_0^{+}}}}+\sqrt{\frac{ r_0^{-}}{f'(r)|_{r_0^{-}}}}\right),
\end{equation}
with $r_0^{\pm}$ being the radius of circular null geodesics for the corresponding mode. Furthermore we have explicitly related the shadow radius to the real part of QNMs in the eikonal regime corresponding to the prograde and retrograde mode, respectively. As a particular example, we have computed the typical black hole shadow radius for some well known black hole solutions including the Kerr black hole, Kerr-Newman black hole and higher dimensional black hole solutions described by the Myers-Perry black hole. 
\end{abstract}
\maketitle

\section{Introduction}

Einstein's general theory of relativity revolutionized how we view time, space and gravity. This theory has important astrophysical implications, perhaps the most interesting prediction is the existence of BHs. The strongest evidence to date are the recent experimental announcements the detection of
gravitational waves (GWs) \cite{AbbottBH} by
the LIGO and VIRGO observatories and the captured image of the black hole shadow of a
supermassive M87 black hole by the Event Horizon Telescope collaboration \cite%
{Akiyama1,Akiyama4}.  However there are many other astrophysical phenomena inferring  their existence for example the high-energy phenomena such as X-ray emission and jets, and the motions of nearby objects in orbit around the hidden mass.

Gravity waves and BH shadows have opened a new window in our understanding of the Universe. Based on observations, in the near future, we can test many of alternative theories of gravity. In addition, using high-resolution images of BH shadows  we can measure the angular diameter the BH or detect tiny effects which are out of reach of the present technology. Thus it remains
an open question if future astronomical observations can potentially detect such important effects.  Usually there are three stages when we study the evolution of binary black holes. The first stage is the so called inspiral stage and has an inspiral phase signal which encodes valuable information about the masses and the spins of compact objects treated by the post-Newtonian
approximation \cite{Blanchet}. The second stage is the merger phase and describes
the rapid collapse, say of two BHs to form a bigger BH \cite{Pretorius,Campanelli,Baker}. Finally there is the
ringdown phase is the final stage and describes a perturbed BH that emits GWs in the form of
quasinormal radiation \cite{BertiCardosoWill}. The perturbation theory of Schwarzschild black hole and its stability under small perturbations was studied in Refs. \cite{Regge,Zerilli}. Since then the perturbations or also known as
the quasinormal modes (QNMs) have been investigated by using other analytic and numerical methods \cite{1,2,3,4,5,6,7,8,9,10,11,12,13,14,15}. 

On the other hand, the shadow of the Schwarzschild BH was first studied by Synge \cite{Synge66} and Luminet \cite{Luminet79} then subsequently the Kerr BH was studied by Bardeen \cite{DeWitt73}.  At first it seems that there is no direct connection between QNMs and shadow radius. However, it turns out that such a connection in fact exists. To understand this connection we fist note that Cardoso et al. \cite{cardoso}  (but see also Refs. \cite{Hod:2017xkz,Wei:2019jve}) where the real part of the QNMs is related to the angular velocity of the last circular null geodesic. Later on Stefanov et al. \cite{Stefanov:2010xz}  found a connection between black-hole quasinormal modes in the eikonal limit and lensing in the strong deflection limit. In a recent work (see \cite{Jusufi:2019ltj}), we have argued that is more suitable to  relate the real part of the QNMs with the shadow radius in spherically symmetric and static black holes.
As we know the rotation is important in studying the shape of the back hole shadow. In fact, mostly of the BHs in the galactic center are expected to rotate, hence determining the angular momentum is of crucial importance. In this paper, we would like to extend the connection between the shadow radius and the real part of the QNMs  for a rotating and asymptotically flat black holes. 

This paper is organized as follows. In Section II, we review the shadow of rotating BHs obtained via Newman--Janis algorithm. In Sec. III we will present a method to compute the typical shadow radius based on the correspondence between the shadow radius and the real part of QNMs in the eikonal regime.  In Secs. IV-VI, we apply this method to obtain the shadow radius for Kerr BH, KNBH and MP BH. In Sec. VII, we briefly discuss the Teukolsky equation. Finally in Sec. VIII, we comment on our results.

\section{Shadow of rotating black holes}
Consider the rotating black hole spacetime given by the metric \cite{Azreg-Ainou:2014pra}
\begin{align}\label{metric}
&ds^2=\left(1-\frac{2\Upsilon(r) r}{\Sigma}\right)dt^2 +2 a\sin^2\theta \frac{2\Upsilon(r) r}{\Sigma}dt d\phi\\
&-\frac{\Sigma}{\Delta}dr^2-\Sigma d\theta^2 - \frac{[(r^2+a^2)^2-a^2\Delta \sin^2\theta] \sin^2\theta}{\Sigma} d\phi^2,\nonumber
\end{align}
where
 \begin{equation}
\Upsilon(r)=\frac{r (1-f(r))}{2},
\end{equation}
along with
\begin{eqnarray}
 \Delta &= & r^{2}f(r)+a^2,\\
 \Sigma &=& r^{2}+a^{2}\cos^{2}\theta.
\end{eqnarray}
The function $f(r)$ in the metric encodes information about the spacetime geometry of a particular solution.
In order to find the contour of a black hole shadow  we need to separate the null geodesic equations in the general rotating spacetime metric (1) using the Hamilton-Jacobi equation given by
\begin{equation}
\frac{\partial \mathcal{S}}{\partial \sigma}=-\frac{1}{2}g^{\mu\nu}\frac{\partial \mathcal{S}}{\partial x^\mu}\frac{\partial \mathcal{S}}{\partial x^\nu},
\label{eq:HJE}
\end{equation}
where $\sigma$ is the affine parameter, $\mathcal{S}$ is the Jacobi action. For this purpose we can express the action in terms of known constants of the motion as follows
\begin{equation}
\mathcal{S}=-\frac{1}{2}\mu ^2 \sigma + E t - J \phi + \mathcal{S}_{r}(r)+\mathcal{S}_{\theta}(\theta),
\label{eq:action_ansatz}
\end{equation}
where $\mu$ is the mass of the test particle, $E=p_t$ is the conserved energy and $J=-p_\phi$ is the conserved angular momentum. After we take $\mu=0$ one can obtain the following equations of motion 
\begin{eqnarray}\notag
 \Sigma\frac{dt}{d\lambda}&=&a(J-aE\sin^{2}\theta)
       +\frac{r^{2}+a^{2}}{\Delta}\left[E\,(r^{2}+a^{2})-a\,J\right],\\\notag
 \Sigma \frac{dr}{d\lambda}&=& \pm \sqrt{\Re},\\\notag
 \Sigma \frac{d\theta}{d\lambda}&=&\pm \sqrt{\Theta},\\
\Sigma\frac{d\phi}{d\lambda}
     &=&(J\csc^{2}\theta-a\,E)+\frac{a}{\Delta}\left[E(r^{2}+a^{2})-a\,J\right],
\end{eqnarray}
where $\lambda$ is the affine parameter, $J$ is the angular momentum of the photon, $E$ is the energy of the photon and $\mathcal{K}$ is the Carter constant. In addition we have introduced
\begin{eqnarray}
 {\Re}&=&\left(a^2 \,E-a\,J+E\, r^2\right)^2-\Delta
   \left[\mathcal{K}+(J-a\, E)^2\right],\\
 \Theta&=&\mathcal{K}-(J\csc\theta-a\,
   E \sin\theta)^2+(J-a\,E)^2.
\end{eqnarray}

The size and shape the black hole shadow is determined by the unstable circular photon orbits satisfying the following conditions
\begin{equation}\lb{condition}
\Re(r)=0,\;\; \frac{d\Re(r)}{dr} =0 ,\;\;\; \frac{d^2 \Re(r)}{dr^2} >0.
\end{equation}
By using this condition the circular orbit radius $r_{ph}$ of the photon can be obtained and the parameters $\xi\equiv J/E$ and $\eta\equiv\mathcal{K}/E^{2}$ can thus be expressed as
where $X(r)=(r^2+a^2)$, and $\Delta(r)$ is defined by Eq. (3), while $\mathcal{K}$ is known as the Carter separation constant. From these conditions one can show that the motion of the photon can be determined by the following two impact parameters  (see \cite{100})
\begin{equation}
\xi=\frac{X_{ph}\Delta'_{ph}-2\Delta_{ph}X'_{ph}}{a\Delta'_{ph}},
\label{eq:xi}
\end{equation}
\begin{equation}
\eta=\frac{4a^2X'^2_{ph}\Delta_{ph}-\left[\left(X_{ph}-a^2\right)\Delta'_{ph}-2X'_{ph}\Delta_{ph} \right]^2}{a^2\Delta'^2_{ph}}.
\label{eq:eta}
\end{equation}

\begin{figure}
\includegraphics[width=7.4cm]{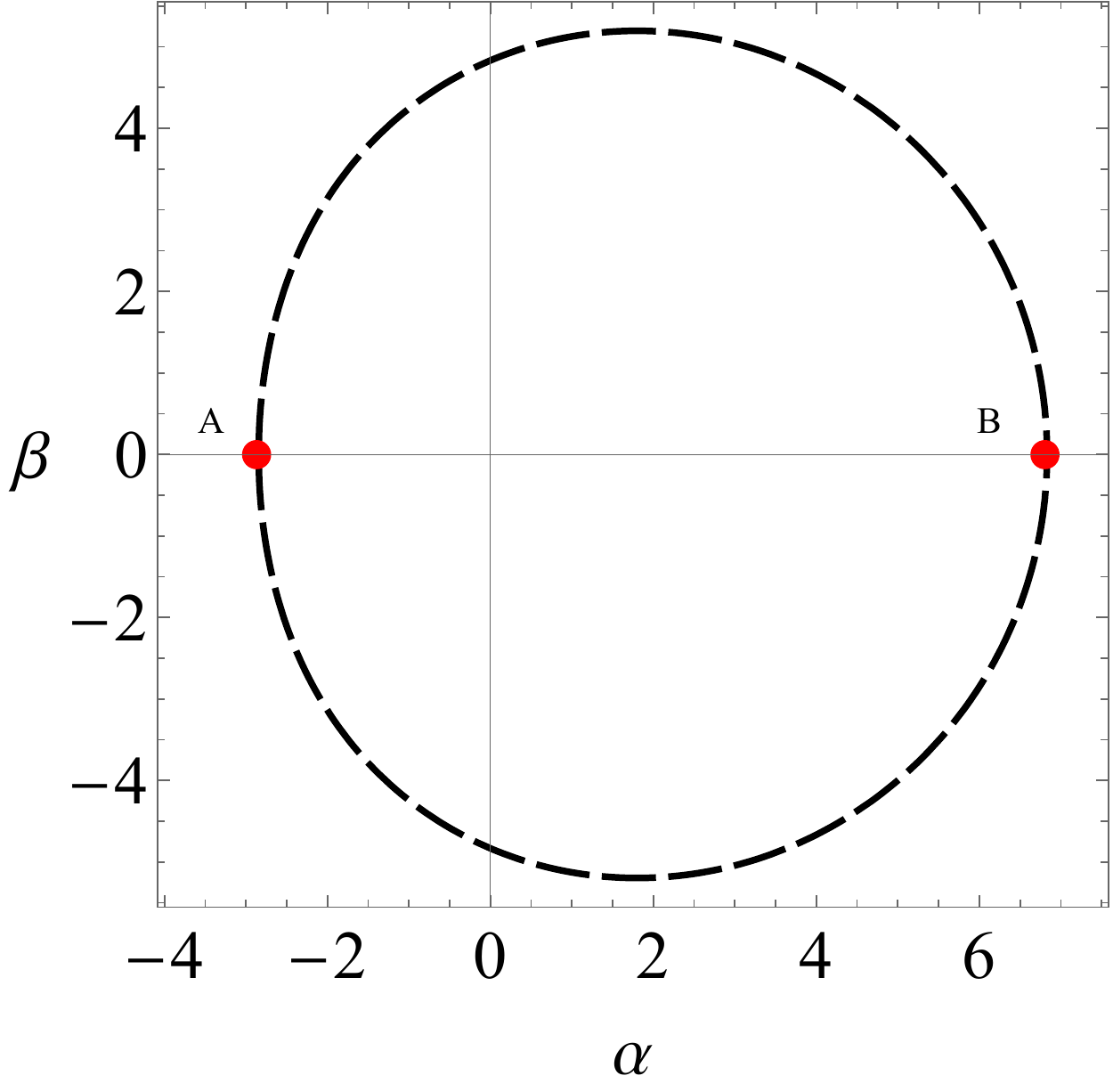}
\caption{Shadow of Kerr black hole with $M=1$, angular momentum $a=0.9$, and an inclination angle $\theta_{0}=\pi/2$.}
\end{figure}
One constraint for the value of the photon's circular orbit radius is ${\Re(r_{ph})>0}$.
The shape of the shadow seen by an observer at spatial infinity can be obtained from the geodesics of the photons and described by the celestial coordinates
\begin{equation}\label{celkerrone}
\alpha=-\xi\csc\theta_{0},
\end{equation}
\begin{equation}\label{celkerrtwo}
 \beta=\pm\sqrt{\eta+a^{2}\cos^{2}\theta_{0}-\xi^{2}\cot^{2}\theta_{0}},
\end{equation}
where $\theta_{0}$ is the inclination angle of the observer. Working in the case with $\theta_0=\pi/2$, in the present paper, we shall use the  definition adopted in Refs. \cite{Zhang:2019glo,Feng:2019zzn} where the typical shadow radius is defined in terms of the leftmost and rightmost coordinates 
\begin{equation}
\bar{R}_S=\frac{1}{2}\left(\alpha^{+}-\alpha^{-}\right),
\end{equation}
along with the condition $\beta(r=r_A)=\beta(r=r_B)=0$. The typical shadow radius is of major interest since it represents an observable quantity.  On the other hand, there is no unique way to define this quantity as a result different definitions have been used (see \cite{000,01,00,111,22,33,44,55,66,77,88,99,100,1111,222,333,444,Banerjee:2019nnj}). 
\begin{figure*}
\includegraphics[width=7.4cm]{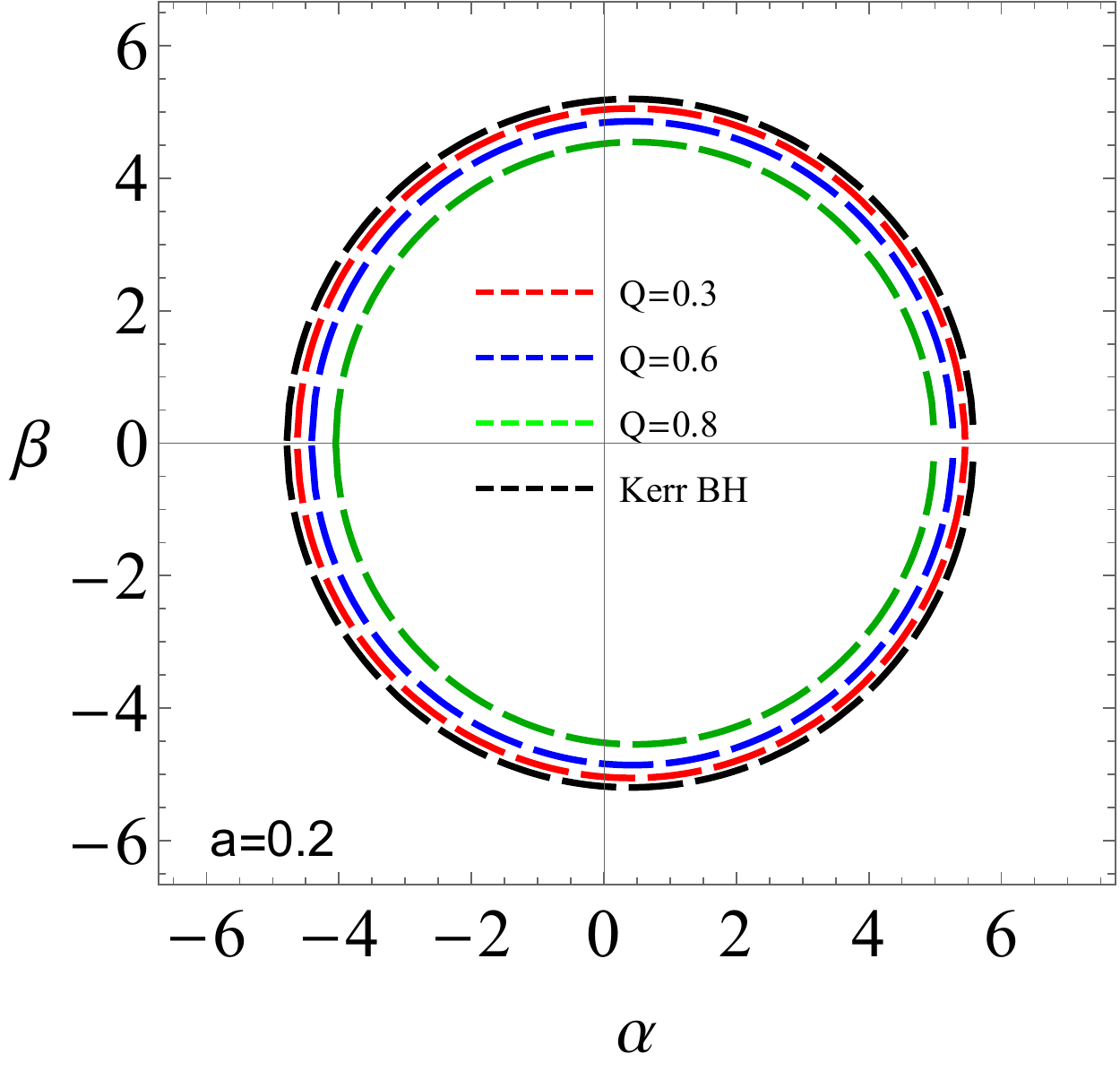}
\includegraphics[width=7.4cm]{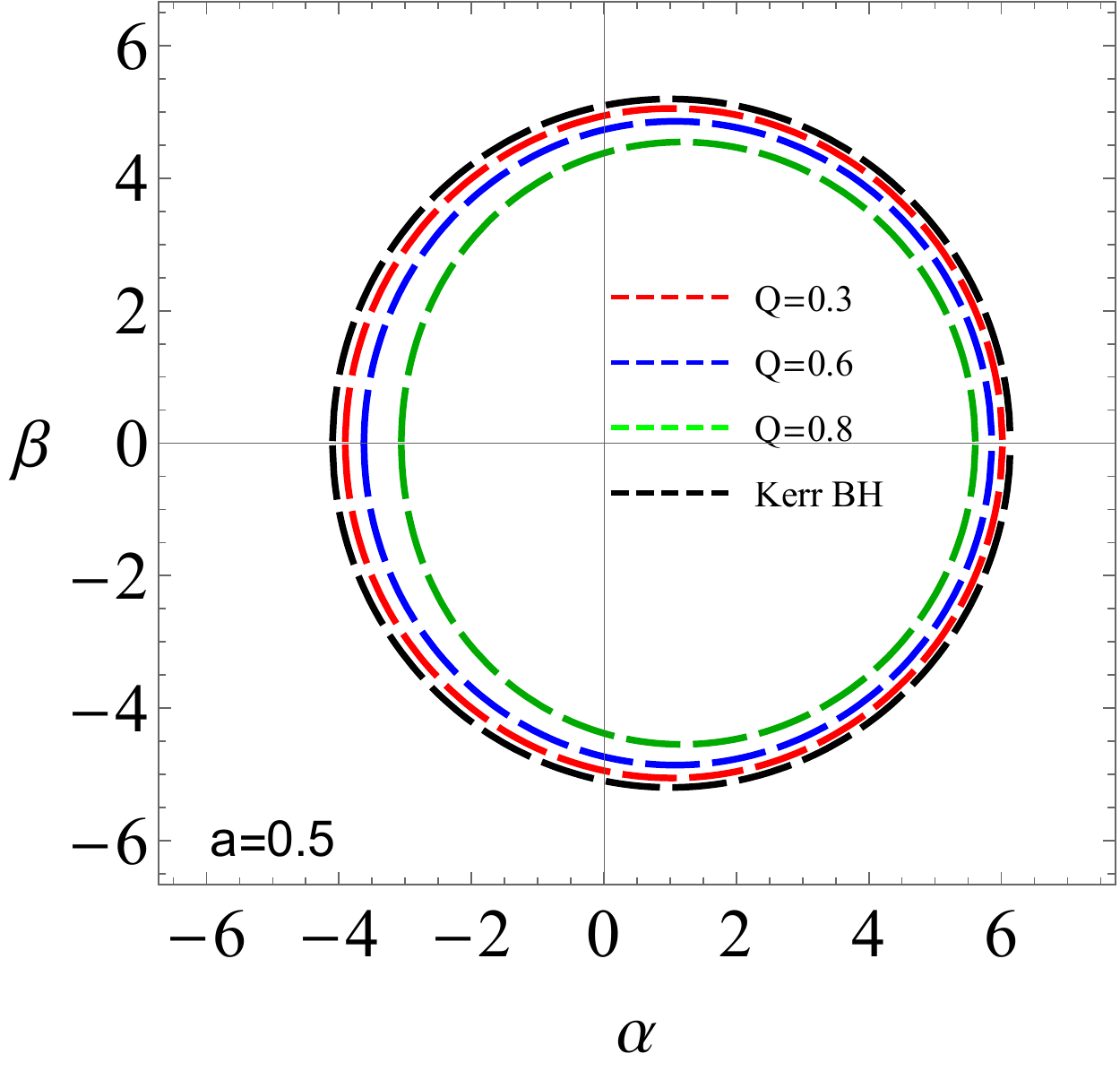}
\caption{The typical shadow radius for the Kerr-Newman black hole for different values of electric charges and angular momentum. We clearly see that increasing the electric charge the shadow radius shrinks. }
\end{figure*}

\section{Connection between shadow radius and QNMs}
QNMs are characteristic modes which encode important information about the stability of the black hole under small perturbations. In order to study these characteristic modes one must impose an outgoing boundary condition at infinity and an ingoing boundary condition at the horizon. In general QNMs can be written in terms of the real part and the imaginary part representing the decaying modes
\begin{equation}
\omega_{QNM}=\omega_{\Re}-i \omega_{\Im}.
\end{equation}

In a seminal paper by Cardoso et al. \cite{cardoso} it was shown that the real part of the QNMs in the eikonal limit is in fact connected with the angular velocity of the last null geodesic. Furthermore the imaginary part was shown to be related with the Lyapunov exponent \cite{cardoso}
\begin{equation}
\omega_{QNM}=\Omega_c l -i \left(n+\frac{1}{2}\right)|\lambda|.
\end{equation}

Based on these equation Stefanov et al. \cite{Stefanov:2010xz} showed  a link between the QNMs and the strong deflection limit. In particular they found that
\begin{equation}
\Omega_c=\frac{1}{\theta D_{OL}},\,\,\,\lambda=\frac{\ln \tilde{r}}{2 \pi \theta D_{OL}},
\end{equation}
in which $D_{OL}$ represents the distance between the observer and the lens,  $\theta$ gives the angular position of the image that is closest to the BH and finally $\lambda$ is the so-called Lyapunov exponent and determines the instability time scale.  In Ref. \cite{Jusufi:2019ltj} Jusufi showed a connection between the shadow radius and the real part of QNMs. It is straightforward to see this connection by adopting the definition
\begin{equation}\lb{thetas}
\theta=\frac{R_s}{D},
\end{equation}
hence from (17) and (18) we can explicitly relate the real part of the QNMs with the shadow radius  \cite{Jusufi:2019ltj}
\begin{equation}
R_s = \lim_{l \gg 1} \frac{l}{\omega_{\Re}}.
\end{equation}

We point out again that this relation is  accurate only in the eikonal limit having large values of $l$. But one can still use this relation in some cases even for small $l$, for example to investigate the relationship between the QNMs and shadow radius upon different physical quantities (see, \cite{Liu:2020ola}) due to their inverse relation. Notice that an equivalent expression in terms of QNMs and angular velocity was written in Refs. \cite{Hod:2017xkz,Wei:2019jve} but no explicit connection whatsoever between the shadow radius and QNMs was mentioned in those papers. From the physical point of view, it is therefore interesting to use Eq. (20) in studying various problems relating the gravity waves and black hole shadow hole. Equation (20) simply reflects the fact that for a distant observer the gravitational waves can be treated as scalar massless particles propagating along the last null unstable and slowly leaking out to infinity.  We point out here that this correspondence is not guaranteed for gravitational fields, for example in the Einstein-Lovelock theory even in the eikonal limit this correspondence may be violated (see, \cite{Konoplya:2017wot}). 

In what follows, we shall argue a connection between QNMs and shadow radius for the rotating black holes metric. Let us start by writing the rotating spacetime
\begin{equation}
ds^2=g_{tt}dt^2+2g_{t\phi}dt d\phi+g_{rr}dr^2+g_{\theta \theta}d\theta^2+g_{\phi \phi}d\phi^2,
\end{equation}
and let us restrict attention to orbits in the equatorial plane $\theta=\pi/2$ for which the appropriate Lagrangian is
\begin{equation}
\mathcal{L}=\frac{1}{2}\left(g_{tt}\dot{t}^2+g_{rr} \dot{r}^2+2 g_{t \phi} \dot{t} \dot{\phi}+g_{\phi \phi}\dot{\phi}^2\right).
\end{equation}

The generalized momenta following from this Lagrangian are
\begin{eqnarray}
p_t&=&g_{tt}\dot{t}+g_{t \phi} \dot{\phi}=E\\
p_{\phi}&=&g_{t\phi}\dot{t}+g_{\phi \phi} \dot{\phi}=-J\\
p_r&=&g_{rr}\dot{r}
\end{eqnarray}

From the above equations it is easy to obtain
\begin{equation}
\dot{\phi}=\frac{g_{t \phi} E+g_{tt} J}{g_{t \phi}^2-g_{tt}g_{\phi \phi}},
\end{equation}
and 
\begin{equation}
\dot{t}=-\frac{g_{\phi \phi} E+g_{t \phi} J}{g_{t \phi}^2-g_{tt}g_{\phi \phi}}.
\end{equation}

Using the Hamiltonian 
\begin{equation}
\mathcal{H}=p_t \dot{t}+p_{\phi }\dot{\phi}+p_{r} \dot{r}-\mathcal{L},
\end{equation} 
and considering the null geodesics we obtain
\begin{equation}
r^2 \mathcal{V}_r=r^2 E^2+a^2 E^2-J^2+(aE-J)^2(1-f(r)),
\end{equation}
where we have used $\mathcal{V}_r=\dot{r}^2$. The conditions for the existence of circular geodesics can be written as
\begin{equation}
\label{Vreq}
\mathcal{V}_r=\mathcal{V}'_r=0,
\end{equation}
thus we obtain the following relations
\begin{equation}
r^2 E^2+a^2 E^2-J^2+(aE-J)^2(1-f(r))=0,
\end{equation}
and
\begin{equation}
2r E^2-(aE-J)^2f'(r)=0.
\end{equation}

Now the key point is to use the geometric-optics correspondence between the parameters of a quasinormal mode, and the conserved quantities
along geodesics \cite{Yang:2012he}. In particular, the energy of the particle can be identified with the real part of QNMs, hence we can identify
\begin{equation}
E \to \omega_{\Re},
\end{equation}
and the azimuthal quantum number corresponds to angular momentum
\begin{equation}
J \to m.
\end{equation} 

In the eikonal limit in the rotating spacetimes we have
\begin{equation}
m=\pm l,
\end{equation}
corresponding to the prograde and retrograde modes, respectively. With these identifications  we can therefore write Eq. (20) as follows
\begin{equation}
\omega_{\Re}^{\pm} = \lim_{l \gg 1}  \frac{m}{R_S^{\pm}}.
\end{equation}

Now let us introduce the following quantity
\begin{equation}
R_s=\frac{J}{E}
\end{equation}
and then combine this relation with Eq. (32), yields 
\begin{equation}
2r-(a-R_s)^2\,f'(r)=0,
\end{equation}
which has the solution
\begin{equation}
R_s^{\pm}=a \pm \sqrt{\frac{2 r_0^{\pm}}{f'(r)|_{r_0^{\pm}}}}.
\end{equation}
In the last equation we have evaluated $r$ at the point $r=r_0$, which gives the radius of circular null geodesics.
With this result in hand, from Eq. (31) we obtain
\begin{equation}
r_0^2-\frac{2 r_0}{f'(r)|_{r_0^{\pm}}}f(r_0)\mp 2 a \sqrt{\frac{2 r_0}{f'(r)|_{r_0^{\pm}}}}=0.
\end{equation}

Our aim is to compute the shadow radius, however in general the shape of the shadow depends on the observer’s viewing angle $\theta_0$. In the case with $\theta_0=\pi/2$ we can adopt the definition (15) known as the typical shadow radius  which can be written as 
\begin{equation}
\bar{R}_s=\frac{1}{2}\left(R^+_s|_{r_0^+}-R^-_s|_{r_0^-}\right)
\end{equation}
where $r_0^{\pm}$ is determined from Eq. (40).  This relation simply follows if we combine relation (13) along with the definitions (15) and (37) and choosing the inclination angle $\pi/2$.  Now let us use the quantity given by  Eq. (41) then, a simple algebra from the last equation, results with an equation for the typical shadow radius
\begin{equation}
\bar{R}_s=\frac{\sqrt{2}}{2}\left(\sqrt{\frac{ r_0^{+}}{f'(r)|_{r_0^{+}}}}+\sqrt{\frac{ r_0^{-}}{f'(r)|_{r_0^{-}}}}\right).
\end{equation}

To the best of our knowledge, this is new equation and has not been reported before. It is important to mention that the roots of the above equation $r_0^{\pm}$ must be chosen that both are outside of the horizon. Finally, sometimes it may be useful to evaluate QNMs numerically and to obtain the shadow radius in terms of the real part of QNMs. For this purpose we can express the typical shadow radius in terms of QNMs as follows
\begin{equation}
\bar{R}_s=\lim_{l>>1}  \frac{m}{2}\left(\frac{1}{\omega^{+}_{\Re}|_{r_0^+} }- \frac{1}{\omega^{-}_{\Re}|_{r_0^-}}\right),
\end{equation}
provided $m=l$. Alternatively we can set $m=-l$ and chose an appropriate definition to obtain the same result. As a limiting case we can consider the static spacetime when $a=0$ implying $\omega^{+}_{\Re}=-\omega^{-}_{\Re}=\omega_{\Re}$, thus we obtain Eq. (20).

\section{Kerr black hole}
In this section we are going to calculate the typical shadow radius for the Kerr black hole in the equatorial plane having 
\begin{equation}
\Delta=r^2f(r)+a^2=r^2-2Mr+a^2.
\end{equation}
Note that $M$ is the mass of the black hole and $a$ is the rotation parameter defined by $a\equiv J/M$ with $J$ the angular momentum of the black hole. 
Using the relation (42) we find
\begin{equation}
\bar{R}_s=\frac{1}{2}\left(r_0^+ \sqrt{\frac{ r_0^{+}}{ M}}+r_0^-\sqrt{\frac{ r_0^{-}}{M}}\right).
\end{equation}
where $r_0^{\pm}$ corresponds to the prograde and retrograde mode and are determined by solving the equation 
\begin{equation}
3 r_0 M -r_0^2 \pm 2 a    \sqrt{r_0 M}=0.
\end{equation}
\begin{table}[tbp]
\begin{tabular}{|l|l|l|l|l|l|}
\hline
\multicolumn{1}{|c|}{ } &  \multicolumn{1}{c|}{  $m=100$ } & \multicolumn{1}{c|}{  $m=100$ } & \multicolumn{1}{c|}{Kerr}\\\hline
  $a$ &\,\,\,\,$\omega_{\Re}^{-}$  &\,\,\,\,$\omega_{\Re}^{+}$  & \,\,\,\,$\bar{R}_s$   \\ \hline
0.0 & -19.24500897 & 19.24500897   & \,\,\,\,\,$3 \sqrt{3}$ \\ 
0.1 & -20.86329192 & 17.87819737   & 5.193256265 \\ 
0.2 & -22.81414428 & 16.70659753  & 5.184452475  \\
0.3 & -25.21936783 & 15.68985638 &  5.169375540  \\ 
0.4 & -28.27162422 & 14.79821770 &  5.147343015 \\ 
0.5 & -32.29695986 & 14.00922085 &  5.117211190 \\ 
0.6 & -37.90041288 & 13.30557023 &  5.077071505 \\ 
0.7 & -46.36584147 & 12.67371538 &  5.023553105 \\  
0.8 & -61.07624270 & 12.10287853 &  4.949897520 \\
0.9 & -95.74679312 & 11.58437233 &   4.838370314 \\\hline
\end{tabular}
\caption{The Kerr shadow radius against the angular momentum. }
\end{table}
\begin{figure}
\includegraphics[width=8.4cm]{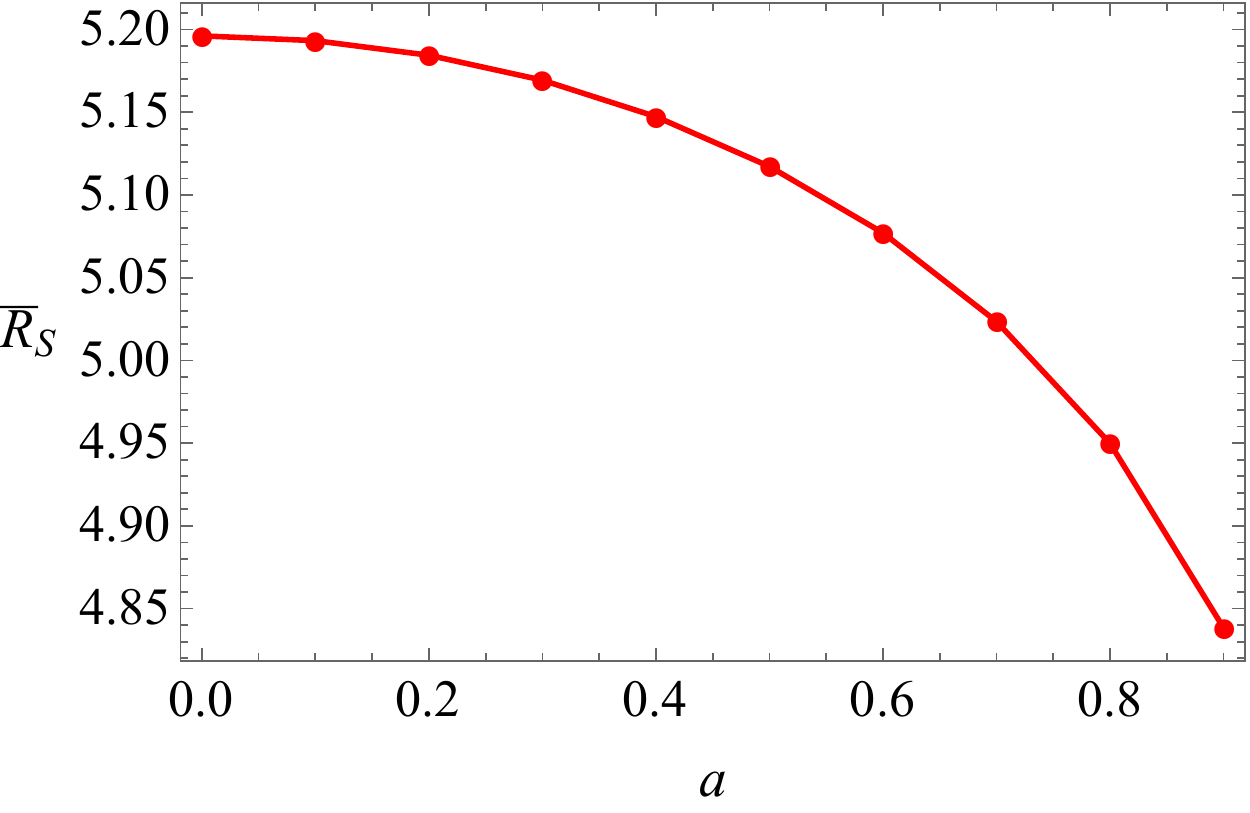}
\caption{The typical shadow radius for the Kerr black hole against the angular momentum. }
\end{figure}

In Table I we show the values for the typical shadow radius using Eq. (42) for different values of the angular momentum $a$. In addition we have evaluated the real part of QNM for $m=l=100$ using 
\begin{equation}
\omega_{\Re}^{\pm}=\lim_{l>>1}\frac{m}{a \pm \sqrt{\frac{2 r_0^{\pm}}{f'(r)|_{r_0^{\pm}}}}}.
\end{equation}

The last equation is nothing but the one obtained in Ref. \cite{2} in the case of Kerr-Newman black hole. From Fig. 3, we observe that the shadow radius decreases with an increase of the angular momentum. Suppose a spinning black hole has an angular momentum $a=0.9$, this means a decrease of the shadow radius $\Delta \bar{R}_s=0.357782110$. We can estimate now the change in the angular radius  $\Delta \theta_s = \Delta R_s M/D$. In the case of M87, for the  supermassive black hole M87 mass we have $M = 6.5 \times  10^{9}$M\textsubscript{\(\odot\)} and $D =16.8$ Mpc is the distance between the Earth and M87 center black hole. We find  $\Delta \theta_s = 9.87098 \times 10^{-6} \Delta R_s(M/$M\textsubscript{\(\odot\)})$(1kpc/ D) \mu as= 1.366416092 \mu as$. In the case of Sgr A$^{*}$ black hole we have $M = 4.3 \times  10^{6}$M\textsubscript{\(\odot\)} and $D =8.3$ kpc yielding a change in the angular radius of the order $\Delta \theta_s =1.829655207 \mu as$.  
As a special case we can obtain the shadow radius for the Schwarzschild black hole. Letting $a \to 0$ one has $r_0=r_0^{+}=r_0^{-}=3M$, yielding
\begin{equation}
\bar{R}_s=r_0\sqrt{\frac{r_0}{M}}=3 \sqrt{3}M.
\end{equation}

Furthermore we can use (42) along with $r_0=2 f(r_0)/f'(r)|_{r_0}$, to rewrite the last equation as follows
\begin{equation}
\bar{R}_s=\frac{r_0}{\sqrt{f(r_0)}}
\end{equation}
which is exactly the same result reported for the static spacetime \cite{Jusufi:2019ltj}.

\section{Kerr-Newman black hole}
In the case of the Kerr-Newman BH according to Eq. (3) we can write
\begin{equation}
\Delta=r^2-2Mr+Q^2+a^2,\,\,\,f(r)=1-\frac{2M}{r}+\frac{Q^2}{r^2}
\end{equation}
Using Eq. (42) it is not difficult to obtain
\begin{equation}
\bar{R}_s=\frac{1}{2}\left(\frac{{(r_0^{+})^{2}}}{\sqrt{Mr_0^+-Q^2}}+\frac{(r_0^{-})^2}{\sqrt{Mr_0^--Q^2}}\right).
\end{equation}
in which the corresponding radius for the circular geodesics can be found after solving the following equation
\begin{equation}
3 r_0 M-2Q^2-r_0^2 \pm  2 a  \sqrt{Mr_0-Q^2}=0.
\end{equation}

For given values of $M$, $Q$ and $a$ we can find the value of $r_0^{\pm}$ corresponding to the prograde and retrograde mode, respectively. In Table II we present our results for the values for the typical shadow radius against the charge for a given value of the angular momentum. We have also evaluated the the real part of QNM for $m=100$.  From Fig. 4, we observe that the shadow radius decreases with an increase of the electric charge and this result is verified in Fig. 2 by means of the geodesic approach. But see also Ref. \cite{444}. 
Finally as a special case when $a=0$ we obtain the shadow radius for the RN black hole with
\begin{equation}
r_0=r_0^+=r_0^-=\frac{1}{2}\left(3M+\sqrt{9 M^2-8 Q^2}    \right)
\end{equation}
and
\begin{equation}
\bar{R}_s=\frac{{r_0^{2}}}{\sqrt{Mr_0-Q^2}}.
\end{equation}

We point out that QNMs in Kerr-Newman spacetime have been studied in Ref. \cite{1,2}. 
\begin{table}[tbp]
\begin{tabular}{|l|l|l|l|l|l|}
\hline
\multicolumn{1}{|c|}{ a=0.2 } &  \multicolumn{1}{c|}{  $m=100$ } & \multicolumn{1}{c|}{  $m=100$ } & \multicolumn{1}{c|}{Kerr-Newman}\\\hline
  $Q$ &\,\,\,\,$\omega_{\Re}^{-}$  &\,\,\,\,$\omega_{\Re}^{+}$  & \,\,\,\,$\bar{R}_s$   \\ \hline
  0.0 & -22.81414428 & 16.70659753 & 5.184452475  \\ 
0.1 & -22.86350943 & 16.72920866  & 5.175675395  \\ 
0.2 & -23.01429916 & 16.79788649  & 5.149127295  \\
0.3 & -23.27508381 & 16.91527287 &  5.104128545  \\ 
0.4 & -23.66197213 & 17.08616391 &  5.039439605 \\ 
0.5 & -24.20220158 & 17.31822837 &  4.953059103 \\ 
0.6 & -24.94164511 & 17.62335128 &  4.841824146 \\ 
0.7 & -25.96181391 & 18.02021646 &  4.700566946 \\  
0.8 & -27.42534331 & 18.53970579 &  4.520045660 \\
0.9 & -29.74270581 & 19.23795448 &   4.280113360 \\\hline
\end{tabular}
\caption{The real parts of QNMs and the shadow radius for different values of the electric charge. }
\end{table}
\begin{figure}
\includegraphics[width=8.4cm]{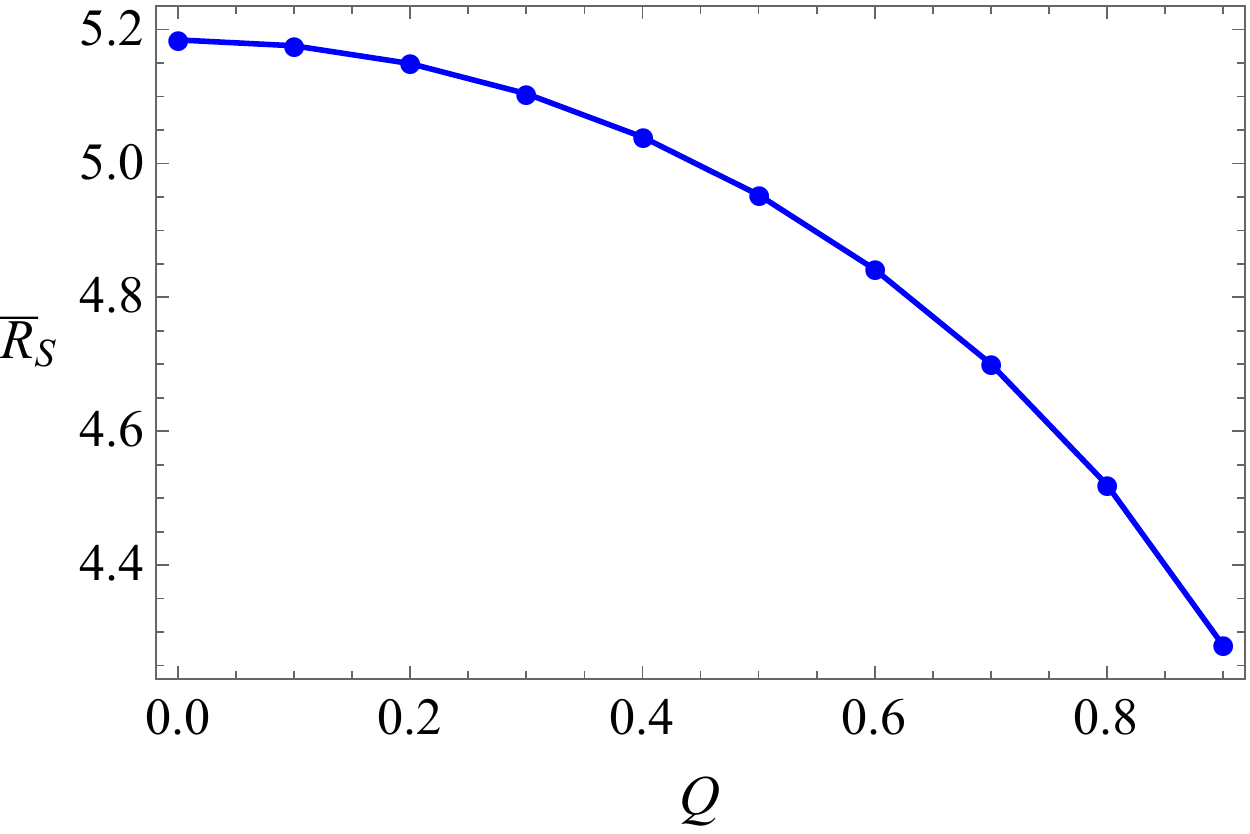}
\caption{The typical shadow radius for the Kerr-Newman black hole against the charge. We have set $M=1$ and $a=0.2$. }
\end{figure}

\section{Myers-Perry black hole}
It is of considerable importance to investigate and extend this method to higher
dimensional black hole solutions. Rotating
black hole solutions in $d$- dimensions are known as
Myers-Perry black holes described by the following metric 
\begin{eqnarray}
 ds^{2}&=&\left(\frac{\Delta-a^2\sin^2\theta}{\rho^{2}}\right)dt^{2}-\frac{\rho^{2}}{\Delta}dr^{2}-\rho^{2}d\theta^{2}\\\notag
&-& \frac{\left[(r^{2}+a^{2})^{2}-\Delta a^{2}\sin^{2}\theta\right]\sin^{2}\theta}{\rho^{2}}d\phi^{2}   \\\notag
&+&\frac{2 a (r^2+a^2-\Delta)\sin^2\theta}{\rho^{2}}dtd\phi -r^2 \cos^2\theta d\Omega^2_{d-4},\label{metric}
 \end{eqnarray}
where $d\Omega^2_{d-4}$ denotes the standard metric of the unit $d-4$ - sphere and
\begin{eqnarray}
 \Delta=r^{2}+a^2-\mu r^{5-d},\,\,\,\,\,
 \rho^{2}=r^{2}+a^{2}\cos^{2}\theta,
\end{eqnarray}

For simplicity, in the present paper we are considering the simplest case having only one angular momentum parameter $a$.  Going through the same steps we can obtain the equation for the radius of circular null geodesics given by \cite{cardoso}
\begin{equation}
\frac{d-1}{d-3}r_0^2 \pm 2 a \sqrt{\frac{2 r_0^{d-1}}{(d-3) \mu }}-\frac{2 r_0^{d-1}}{(d-3)\mu}=0.
\end{equation}

The conditions for the existence of circular geodesics results with the following two equations 
\begin{equation}
r^2+a^2-R_s^2+\mu r^{3-d}(a-R_s)^2=0,
\end{equation}
and
\begin{equation}
2 r+\mu r^{3-d}(a-R_s)^2=0.
\end{equation}
evaluated at $r_0$.
Furthermore we are going to consider the case of $d=5$ since the equation simplify considerably. The radius of the circular null geodesics (57) reduces to  
\begin{equation}
r_0^{\pm}=\sqrt{2}\sqrt{\mu \pm a \sqrt{\mu}}.
\end{equation}

Combining the last result from Eq. (58) we obtain
\begin{equation}
R_s^{\pm}=a \pm \frac{2 (\mu \pm a\sqrt{\mu}) }{\sqrt{\mu}}
\end{equation}

Finally for the typical radius we obtain
\begin{equation}
\bar{R}_s=2 \sqrt{\mu}.
\end{equation}

In other words, we found that the typical shadow radius in the equatorial plane remains unaffected by the rotational parameter $a$. In the case $d=5$ the mass parameter $\mu$ can be written as $\mu=8M/3\pi$. Working in units $\mu=1$, we find $\bar{R}_s=2$. This result is in perfect agreement with Ref. \cite{Amir:2017slq} where authors studied the black hole shadow using geodesic approach. 

\section{Teukolsky equation}
An intuitive geometric correspondence between high-frequency QNMs and angular  frequencies in the Kerr spacetime has been studied in Ref. \cite{Yang:2012he}. One can consider the radial Teukolsky equation in Kerr spacetime using the separation of 
variables (see \cite{Yang:2012he} for details)
\begin{equation}
u(t,r,\theta,\phi) = e^{-i \omega t }e^{i m \phi}u_r(r)u_{\theta}(\theta) \, .
\end{equation}
Now at the relevant order in $l \gg 1$, the angular equation for 
$u_\theta(\theta)$ can be stated as follows \cite{Yang:2012he}
\begin{align}
\frac{1}{\sin \theta}\frac{d}{d\theta}\left[\sin{\theta}
\frac{d u_{\theta}}{d \theta}\right]+
\left[a^2\omega^2\cos^2{\theta}-\frac{m^2}{\sin^2{\theta}}+A_{lm}\right]
u_{\theta}=0 \,,
\end{align}
with $A_{lm}$ being the angular eigenvalue of this equation. 
The radial equation $u_r(r)$ reduces to \cite{Yang:2012he}
\begin{equation}
\frac{d^2 u_r}{dr_*^2} +V^r u_r=0.
\end{equation}
in which
\begin{equation}
V^r=\frac{[\omega(r^2+a^2)- ma ]^2 -\Delta  \left[A_{lm}(a \omega) +a^2\omega^2 -2 m a \omega\right] }{(r^2+a^2)^2}.
\end{equation} 

It was shown in Ref. \cite{Yang:2012he} that  at the leading and 
next-to-leading order one can find $\omega_{\Re}$ by using the condition
\begin{equation}
V^r(r_0,\omega_{\Re})=\left.\frac{\partial V^r}{\partial r}\right|_{(r_0,\omega_{\Re})}
=0 \, 
\end{equation}
yielding
\begin{equation}
\lim_{l>>1}\Omega_{\Re}^{0} =\frac{a(r_{0}-M)\,\mu} {(3M-r_{0})r_{0}^{2}-a^{2}(M+r_{0})}\,\,,
\end{equation}
where $\mu$ and $\Omega_{{\Re}}$ are defined as follows
\begin{equation}
\mu=\frac{m}{L}\,\,,
\end{equation}
and
\begin{equation}
\Omega_{{\Re}}=\frac{\omega_{\Re}}{L}\,\,,
\end{equation}
in which $L=l+1/2$. Working in the eikonal limit we can therefore use $\mu=\pm1$, since $\lim_{l>>1}L=l$. Making use of the inverse relation 
between the orbital frequency and shadow radius  $\Omega_{\Re}=1/R_s$  to finally obtain 
\begin{equation}
\omega_{\Re}=\lim_{l>>1} \frac{m}{R_s}.
\end{equation}
This is an alternative way of seeing the same problem.
One can use the last equation along with (15) and (68) to find the typical shadow radius. This is precisely what was done in Ref. \cite{Feng:2019zzn}. In the present work, however,  we have explicitly related the shadow radius and the real part of QNMs using the geometric-optics correspondence and the conserved quantities along geodesics.

\section{Conclusion}
In this paper we have shown a connection between the shadow radius and the real part of QNMs in rotating BH spacetimes. This connection is obtained using the correspondence between the parameters of a quasinormal mode and the conserved quantities
along geodesics. For the typical shadow radius we have obtained an equation given by Eq. (42) provided the observer’s viewing angle is $\theta_0=\pi/2$. Alternatively, we have argued that one can express this result in terms of the real part of QNMs given by Eq. (43). We have applied these equations to explore the typical black hole shadow radius for the the Kerr black hole, Kerr-Newman black hole and five dimensional Myers-Perry black hole.  Our results are consistent with the ones obtained via standard methods solving the geodesic equations. 

\section*{Acknowledgments}
The author is grateful to the editor and anonymous referees for their valuable comments and suggestions to improve
the paper.

 \end{document}